\documentstyle[12pt]{article}

\begin{document}

\begin{titlepage}

\title{Abelian Sandpile Model on the Husimi Lattice of Square
      Plaquettes.}

\author{Vl.V.~Papoyan\thanks{E-mail: papoyanv@theor.jinrc.dubna.su} \\[2mm]
and \\[2mm]
R.R.~Shcherbakov\thanks{E-mail: shcher@thsun1.jinr.dubna.su}
\thanks{Permanent address: Department of Physics, Yerevan State
University, Alek Manoukian St.1, Yerevan, 375049, Armenia.}\\[2mm]
{\small \sl Bogoliubov Laboratory of Theoretical Physics,} \\
{\small \sl JINR, 141980 Dubna, Russia.}
}

\date{}

\maketitle

\begin{abstract}
An Abelian sandpile model is considered on the Husimi lattice of
square plaquettes.
Exact expressions for the distribution of height probabilities
in the Self-Organized Critical state are derived.
The two-point correlation function for the sites deep inside the Husimi
lattice is calculated exactly.
\end{abstract}

\thispagestyle{empty}
\end{titlepage}

\newpage
\section{Introduction.}

In recent years, there has been considerable interest
in different dynamical models which can evolve without any
tuning of parameters to the {\it Self-Organized Critical}
(SOC) state.

The concept of Self-Organized Criticality has been introduced
by Bak, Tang, and Wiesenfeld~\cite{BTW}, through simple
cellular automaton models known as {\it sandpiles}, to explain
the temporal and spatial scaling in dynamical dissipative
systems.

Later on, Dhar and Majumdar, in a number of articles~[2-4], have studied
the so-called {\it Abelian sandpile models} (ASM's) for
they show a nontrivial analytically tractable example of the SOC
behavior. For these models some ensemble-average quantities
have been calculated on the square lattice~[2,4-7].

In this article, we consider ASM on the Husimi lattice of
square plaquettes. One of the remarkable features of the Husimi-
or Bethe-like lattices is the exact solvability of different
spin, gauge and dynamical models defined on them~[8-14].
One might also hope, by examining the form of exact
solutions for these lattices which allow only a limited number
of closed configurations, to predict a general behavior for
the lattice models of principal interest.

It is well known that the exact solutions obtained for
different spin models on the Bethe lattice are equivalent
to the Bethe-Peierls approximation and they give a more precise
description of the critical behavior of these models than
the mean-field approximation technique~\cite{E}. In turn, the Husimi
lattice can be considered as a next step in this list of
approximations and, as it has been shown by Monroe~\cite{Mon}
for the two- and three-site interacting spin systems, this
approximation improves results obtained on the Bethe lattice.
In section {\bf 6} we bring a table where we compare the
distribution of height probabilities in the SOC state
of ASM calculated on different lattices and again
we see that the results obtained on the Husimi lattice are in good
agreement with the known exact results on the square lattice~\cite{MD,P}.
This can be explained by the presence of elementary loops in the
Husimi lattice that reproduce  the local structure of usual lattices
more precisely than the Bethe lattice. Hence, for most of flat
and three-dimensional lattices we can find  suitable Husimi-like
lattices and examine different models on them.

The outline of the article is as follows:
in the next section, we define the lattice and ASM on it.
In section {\bf 3}, we find
recursion relations for the numbers of allowed configurations in
the SOC state.
In section {\bf 4}, we  exactly compute the distribution of height
probabilities.
In section  {\bf 5}, we calculate the two-point correlation function
for the sites deep inside the lattice.
In section  {\bf 6}, we bring some concluding remarks.

\section{Lattice and Model.}

A pure Husimi tree of square plaquettes~\cite{EF} can be
constructed recurrently.
As a basic building block, we will take an elementary plaquette (Fig.1a).
This basic block will be called the {\it first-\-generation
branch}. To construct the {\it second-\-generation branch}, we will
attach a single basic building block at each free site of the
first-\-generation branch except the base site (root) (Fig.1b).
Continuing this process we develop higher-generation branches.
Then, at the final step, we will take four $n$th-\-generation
branches and connect their base sites by the elementary plaquette.
As a result, we will get the graph with the coordination number
$q=4$.

Let as define  ASM on this connected graph of $N$ sites
as follows: to each site $i$ $(1\leq i\leq N)$ we associate
an integer $z_i$ $(1\leq z_i\leq 4)$ which is the height
of a column of sand grains. The evolution of the system is
specified by two rules:

\bigskip
{\bf i.} Addition of a sand grain at a randomly chosen site $i$ increases
$z_i$ by 1.

{\bf ii.} The site $i$ topples if the height $z_i$ exceeds the critical value
$z_c=4$ and sand grains drop on the nearest neighbors.
\bigskip

The number of surface sites of the Husimi tree is comparable
with the interior ones. Hence, the calculation of the thermodynamic
limit of the bulk properties requires special care.
In our work, we define the height distribution of sand grains
and  the two-point correlation function
for the sites deep inside the tree. Using these
interior sites one can construct an infinite lattice,
as they have the same features. Therefore, we will consider
the problem on the Husimi lattice rather than on the Husimi tree.

Any configuration $\{z_i\}$ on the Husimi tree in which $1\leq z_i\leq 4$
is a stable configuration under the toppling rule. These configurations
can be divided into two class: {\it allowed} and {\it forbidden}
configurations~\cite{Dhar}.

In the SOC state, only allowed configurations have a nonzero probabiliy.
Any subconfiguration of heights $F$ on a finite connected set
of sites is forbidden if
$$
z_i\leq q_i, \qquad\qquad \forall\,\, i\, \in \, F\,,
$$
where $q_i$ is a coordination number of a site $i$ in the given
subconfiguration $F$~\cite{Dhar}.

In turn, we can divide the allowed
subconfigurations on an $n$th-generation branch of the Husimi tree
into three nonoverlapping classes:
{\it weakly allowed of the type 1} ($W_1$),
{\it weakly allowed of the type 2} ($W_2$) and
{\it strongly allowed} ($S$) subconfigurations.

Consider an allowed subconfiguration $C$
on the $n$th-generation branch $G_n$ with a root
$a$ (Fig.2a). The coordination number of the root $a$ is $q=2$. Adding
a vertex $b$ to the $G_n$, one defines a
subgraph $G'=G_n \cup b$. If the subconfiguration $C'=C \cup b$ with $z_b=1$
on $G'$ is forbidden, then
$C$ is called the weakly allowed subconfiguration of  type {1} ($W_1$).
Thus, $W_1$ can be locked by one bond, after which it becomes forbidden.

Now add two vertices $b$ and $d$ to $G_n$ and
consider a subconfiguration $C''=C \cup b \cup d$
on  $G''=G_n \cup b  \cup d$ (Fig.2b).
If the subconfiguration $C''=C \cup b \cup d$ with $z_b=1$ and $z_d=1$
on $G''$ is forbidden, then
$C$ is called the weakly allowed subconfiguration of type {2} ($W_2$).

Any allowed subconfigurations defined on the $n$th-generation branches
that cannot be locked by one bond or by two bonds form
a strongly allowed ($S$) class.

It is important to note that any subconfiguration of the $W_1$
type is also of the $W_2$ type. To obtain the nonoverlapping
classes, we always check the subconfiguration first to
belong to the $W_1$ type and only then to the $W_2$ type.

\section{Recursion Relations.}

\setcounter{equation}0

In this section, we endeavor to find recursion relations for the
numbers of allowed configurations on the $n$th-generation branch $G_n$
of the pure Husimi tree.
Consider now $G_n$ with a root vertex $a$ (Fig.3) that consists
of three $(n-1)$th-generation branches $G_{n-1}^{(1)}$, $G_{n-1}^{(2)}$
and $G_{n-1}^{(3)}$ with roots $a_1$, $a_2$ and $a_3$, respectively.
Let $N_{W_1}(G_n,i)$, $N_{W_2}(G_n,i)$
and $N_{S}(G_n,i)$ be the numbers of distinct $W_1$, $W_2$ and $S$
type subconfigurations on $G_n$ with a given height $z_a=i$
at the root $a$.

Let us also introduce
\begin{equation}
\label{f3.1}
N_{W_1}(G_n)=\sum\limits_{i=1}^4N_{W_1}(G_n,i),
\end{equation}

\begin{equation}
\label{f3.2}
N_{W_2}(G_n)=\sum\limits_{i=1}^4N_{W_2}(G_n,i),
\end{equation}

\begin{equation}
\label{f3.3}
N_S(G_n)=\sum\limits_{i=1}^4N_S(G_n,i).
\end{equation}

These numbers can be expressed in terms of the numbers of allowed
subconfigurations on the three $(n-1)$th-generation branches
$G_{n-1}^{(1)}$, $G_{n-1}^{(2)}$ and $G_{n-1}^{(3)}$:
$$
N_{W_1}(G_n)=N_{S}N_{S}N_{S} + N_{W_1}N_{S}N_{S}+
	     N_{W_2}N_{S}N_{S} + N_{S}N_{W_1}N_{S}+
	     N_{W_2}N_{W_1}N_{S} +
$$
\begin{equation}
\label{f3.4}
	     N_{S}N_{W_2}N_{S}+ N_{W_1}N_{W_2}N_{S}+
	     N_{W_2}N_{W_2}N_{S}+ N_{S}N_{S}N_{W_1}+
	     N_{W_1}N_{S}N_{W_1}+
\end{equation}
$$
	     N_{W_2}N_{S}N_{W_1}+ N_{S}N_{W_2}N_{W_1}+
	     N_{W_2}N_{W_2}N_{W_1}+ N_{S}N_{S}N_{W_2}+
	     N_{W_1}N_{S}N_{W_2}+
$$
$$
	     N_{W_2}N_{S}N_{W_2}+ N_{S}N_{W_1}N_{W_2}+
	     N_{W_2}N_{W_1}N_{W_2}+ N_{S}N_{W_2}N_{W_2}+
	     N_{W_1}N_{W_2}N_{W_2}+ N_{W_2}N_{W_2}N_{W_2}\,,
$$

\begin{equation}
\label{f3.5}
N_{W_2}(G_n)=N_{W_1}(G_n)\,,
\end{equation}

$$
N_{S}(G_n) = 2N_{S}N_{S}N_{S} + N_{W_1}N_{S}N_{S}+
	     2N_{W_2}N_{S}N_{S} + 2N_{S}N_{W_1}N_{S}+
	     N_{W_2}N_{W_1}N_{S} +
$$
\begin{equation}
\label{f3.6}
	     2N_{S}N_{W_2}N_{S}+ N_{W_1}N_{W_2}N_{S}+
	     2N_{W_2}N_{W_2}N_{S}+ N_{S}N_{S}N_{W_1}+
	     N_{W_2}N_{S}N_{W_1}+
\end{equation}
$$
	     N_{S}N_{W_2}N_{W_1}+ 2N_{S}N_{S}N_{W_2}+
             N_{W_1}N_{S}N_{W_2}+ 2N_{W_2}N_{S}N_{W_2}+
             N_{S}N_{W_1}N_{W_2}+ 2N_{S}N_{W_2}N_{W_2}\,,
$$

\bigskip

\noindent
where  the first factor in each term of the sum corresponds to the
$G_{n-1}^{(1)}$ branch, the second one to $G_{n-1}^{(2)}$ and
the third one to $G_{n-1}^{(3)}$.

The fact that the numbers of the $W_1$ type subconfigurations and
the $W_2$ type ones are equal to each other is seen
from the straightforward calculation.

\newpage
Let us introduce

\begin{equation}
\label{f3.7}
X=\frac{N_W}{N_S}\,,
\end{equation}
where $N_W\equiv N_{W_1}=N_{W_2}$.

If we consider graphs $G^{(1)}_{n-1}$, $G^{(2)}_{n-1}$
and $G^{(3)}_{n-1}$ to be isomorphic, then
$N(G^{(1)}_{n-1})=N(G^{(2)}_{n-1})=N(G^{(3)}_{n-1})$ and
from~(\ref{f3.4})-(\ref{f3.7}) one obtains the following
recursion relation
\begin{equation}
\label{f3.8}
X(G_n)= \frac{ 1+4X(G_{n-1})+2X^{2}(G_{n-1}) }
             { 2\left[ 1+3X(G_{n-1})\right] }\,.
\end{equation}

The iterative sequence $\{X(G_n)\}$ starts from the seed
$X(G_0)=\frac{1}{2}$ and converges to the stable point
$X^*=\frac{1+\sqrt{5}}{4}$ that characterizes, in the
thermodynamic limit, the ratio of the $W_1$ type
or the $W_2$ type configurations to the strongly allowed ones.

\section{Distribution of Height Probabilities.}

\setcounter{equation}0

One of the main characteristics that describes the SOC state
is the probability $P(i)$ of having the height $z=i$ at a
given site:
\begin{equation}
\label{f4.1}
P(i)=\frac{N(i)}{N_{\mbox{\scriptsize total}}}\,,
\end{equation}
where $N(i)$ is the number of allowed configurations with
a given value $z=i$ $(1\leq i\leq 4)$ and
$N_{\mbox{\scriptsize total}}=\sum^{4}_{i=1}N(i)$ is
the  total number of allowed configurations on the Husimi
lattice.

Consider now a randomly chosen site $O$ deep inside the
Husimi tree (Fig.4). The number $N(i)$ can be expressed via
the numbers of allowed configurations on the six
$n$th-generation branches $G^{(\alpha)}_n$, $\alpha =1,\ldots,6$.

If $i=1$, then each allowed configuration on the branches $G^{(1)}_n$,
$G^{(3)}_n$, $G^{(4)}_n$ and $G^{(6)}_n$ cannot be of the $W_1$ type.
It is also evident that three $W_2$ type configurations cannot
occur on the neighboring branches $G^{(1)}_n$, $G^{(2)}_n$, $G^{(3)}_n$,
or $G^{(4)}_n$, $G^{(5)}_n$, $G^{(6)}_n$, and so on.

Excluding all such forbidden subconfigurations one finds the following
expression for the number $N(1)$ of allowed configurations with $i=1$
at $O$ for isomorphic branches $G^{(\alpha)}_n$, $\alpha =1,\ldots,6$:
\begin{equation}
\label{f4.2}
N(1)=[1+8X+22X^2+24X^3+9X^4]\prod\limits_{\alpha=1}^{6}N_S(G^{(\alpha)})\,.
\end{equation}

Arguing similarly, one can get
\begin{eqnarray}
\label{f4.3}
N(2) & = & [1+12X+50X^2+84X^3+45X^4]\prod\limits_{\alpha=1}^{6}
N_S(G^{(\alpha)})\,,\\
\label{f4.4}
N(3) & = & [1+12X+56X^2+124X^3+119X^4+24X^5]\prod\limits_{\alpha=1}^{6}
N_S(G^{(\alpha)})\,,\\
\label{f4.5}
N(4) & = & [1+12X+56X^2+128X^3+147X^4+72X^5]\prod\limits_{\alpha=1}^{6}
N_S(G^{(\alpha)})\,.
\end{eqnarray}

For the sites far from the surface in the thermodynamic limit
$(n\rightarrow \infty)$ we have $X=\frac{1+\sqrt{5}}{4}$.
Thus, from(~\ref{f4.1})-(\ref{f4.5}) we get
\begin{equation}
\label{f4.6}
P(1)=\frac{5(\sqrt{5}-2)}{16}\,,
P(2)=\frac{10-3\sqrt{5}}{16}\,,
P(3)=\frac{3(4-\sqrt{5})}{16}\,,
P(4)=\frac{4+\sqrt{5}}{16}\,.
\end{equation}

\section{Two-Point Correlation Function.}

\setcounter{equation}0

The two-point correlation function may be defined as the probability
$P_n(i,j)$ of a stable configuration in which two sites
separated by the distance $n$ have heights $i$ and $j$.
To obtain $P_n(i,j)$ explicitly, we consider two
sites $A_1$ and $A_{n+1}$ deep inside the Husimi lattice (Fig.5).
The left-hand side of the lattice beginning from the vertex $A_k$,
$k=1,\ldots,n$, will be denoted as branch or
subtree $G_k$ with the root $A_k$.
This branch, in turn, consists of the subbranches
$G_{k-1}$, $U_k^{(b)}$, $U_{k-1}^{(b)}$.

Following Dhar and Majumdar~\cite{DM}, we can solve
the problem by the transfer matrix technique
based on the fractal structure of the lattice.

The number of allowed configurations $N(i,j)$ on
the Husimi lattice with fixed heights $i$ at $A_1$
and $j$ at $A_{n+1}$ is expressed via ${\bar N}_{\alpha}(G_n)$,
$N_{\alpha}(U^{(t)}_n)$, $N_{\alpha}(U^{(t)}_{n+1})$,
$N_{\alpha}(U_{n+2})$, $N_{\alpha}(U^{(b)}_{n+2})$ and
$N_{\alpha}(U^{(b)}_{n+1})$, where $\alpha=W_1,\, W_2,\, S$.
The bars above ${\bar N}_{\alpha}(G_n)$ denote the constraint
that the height at $A_1$ is fixed at $i$.

The numbers ${\bar N}_S(G_n)$, ${\bar N}_{W_1}(G_n)$ and
${\bar N}_{W_2}(G_n)$,
in turn, can be expressed in terms of the numbers of allowed
configurations on the branches $U^{(b)}_n$, $U^{(b)}_{n-1}$
and $G_{n-1}$ in the matrix form
\begin{equation}
\label{f5.1}
\left(
\begin{array}{c}
{\bar N}_{W_1}(G_n) \\
{\bar N}_{W_2}(G_n) \\
{\bar N}_{S}(G_n)
\end{array}
\right) =
N_S(U_n^{(b)})\cdot N_S(U_{n-1}^{(b)})\cdot {\bf A}\cdot
\left(
\begin{array}{c}
{\bar N}_{W_1}(G_{n-1}) \\
{\bar N}_{W_2}(G_{n-1}) \\
{\bar N}_{S}(G_{n-1})
\end{array}
\right)\,,
\end{equation}
where
\begin{equation}
\label{f5.2}
{\bf A} = \left(
\begin{array}{ccc}
1+3X+X^2 & 1+4X+3X^2 & 1+4X+3X^2 \\
1+3X+X^2 & 1+4X+3X^2 & 1+4X+3X^2 \\
1+2X     & 2+6X      & 2+7X+4X^2
\end{array}
\right)\,.
\end{equation}

In the thermodynamic limit for the sites deep inside the
lattice we have $X=\frac{1+\sqrt{5}}{4}$.

Hence, one can get
\begin{equation}
\label{f5.4}
\left(
\begin{array}{c}
{\bar N}_{W_1}(G_n) \\
{\bar N}_{W_2}(G_n) \\
{\bar N}_{S}(G_n)
\end{array}
\right) =
\prod\limits_{k} N_S(U_k^{(b)})\cdot N_S(U_{k}^{(t)})\cdot
{\bf A}^{n-1}\cdot
\left(
\begin{array}{c}
{N}_{W_1}(G_1,i) \\
{N}_{W_2}(G_1,i) \\
{N}_{S}(G_1,i)
\end{array}
\right)\,,
\end{equation}
where
\begin{equation}
\label{f5.5}
\left(
\begin{array}{c}
{N}_{W_1}(G_1,i) \\
{N}_{W_2}(G_1,i) \\
{N}_{S}(G_1,i)
\end{array}
\right)=
\left(
\begin{array}{c}
{\bar N}_{W_1}(G_1) \\
{\bar N}_{W_2}(G_1) \\
{\bar N}_{S}(G_1)
\end{array}
\right)\,,
\end{equation}
and the product runs over all subbranches of the subtree $G_n$.

To obtain the two-point correlation function, we ought to
divide $N(i,j)$ by the total number of allowed configurations
of the full lattice in the SOC state
\begin{equation}
\label{f5.6}
P_n(i,j)=\frac{N(i,j)}{N_{\mbox{\scriptsize total}}}\,,
\end{equation}
where $N_{\mbox{\scriptsize total}}$ is given by
\begin{equation}
\label{f5.7}
N_{\mbox{\scriptsize total}} = \frac{521+233\sqrt{5}}{2}
N_S(G_n) N_S(U_n^{(t)}) N_S(U_{n+1}^{(t)}) N_S(U_{n+2})
N_S(U_{n+2}^{(b)}) N_S(U_{n+1}^{(b)})
\end{equation}
with
\begin{equation}
\label{f5.8}
N_S(G_n) = \sum\limits_{i=1}^{4} {\bar N}_S(G_n)\,.
\end{equation}

Then, substituting  ${\bar N}_S(G_n)$  from~(\ref{f5.4}) one can
find for $n > 1$
\begin{equation}
\label{f5.9}
N_{\mbox{\scriptsize total}} = \frac{521\,\sqrt{5}+1165}{40}
\left(
18\,{\sqrt{5}}+40+(3+\sqrt{5})({\lambda_3}-{\lambda_2})\right)
\,{{\lambda_3}^{n-1}}\,\prod N_S(U)\,,
\end{equation}
where
\begin{eqnarray}
\label{f5.9a}
{\lambda_1} & = & 0\,, \\
{\lambda_{2,3}} & = & \frac{21+9\sqrt{5}\mp\sqrt{470+210\sqrt{5}}}{4}\,,
\end{eqnarray}
are eigenvalues of the matrix ${\bf A}$
and the product runs over all subbranches which surround
the path from  $A_1$ to $A_{n+1}$.

Thus, after a rather tedious calculation we obtain
an exact expression for $P_n(i,j)$ in the SOC state
\begin{equation}
\label{f5.10}
P_n(i,j)=P(i)P(j) + p_{ij}\left(\frac{\lambda_3}
{\lambda_2}\right)^{-(n-1)}\,,
\qquad n>1\,,
\end{equation}
where $p_{ij}$ are numerical constants
\begin{eqnarray}
p_{11} & = & 8140 + 3640\,{\sqrt{5}} -
     (1282\,{\sqrt{5}} + 2870)(\lambda_3 - \lambda_2)\,, \nonumber \\
p_{12}=p_{21} & = & 50660 + 22656\,{\sqrt{5}} -
   (2762\,{\sqrt{5}} + 6174)(\lambda_3 - \lambda_2)\,, \nonumber \\
p_{13}=p_{31} & = & 122760 + 54900\,{\sqrt{5}} -
     (3234\,{\sqrt{5}} + 7230)(\lambda_3 - \lambda_2)\,, \nonumber \\
p_{14}=p_{41} & = & 171560 + 76724\,{\sqrt{5}} -
     (3026\,{\sqrt{5}} + 6766)(\lambda_3 - \lambda_2)\,, \nonumber \\
p_{22} & = & 175612 + 78536\,{\sqrt{5}} -
   (\frac{33498}{\sqrt{5}} + 14982)(\lambda_3 - \lambda_2)\,, \nonumber \\
p_{23}=p_{32} & = & 333040 + 148940\,{\sqrt{5}} -
   (9290\,{\sqrt{5}} + 20774)(\lambda_3 - \lambda_2)\,, \nonumber \\
p_{24}=p_{42} & = & 425488 + 190284\,{\sqrt{5}} -
   (\frac{49922}{\sqrt{5}} + 22326)(\lambda_3 - \lambda_2)\,, \nonumber \\
p_{33} & = & 521612 + 233272\,{\sqrt{5}} -
   (\frac{76682}{\sqrt{5}} +34294)(\lambda_3 - \lambda_2)\,, \nonumber \\
p_{34}=p_{43} & = & 605724 + 270888\,{\sqrt{5}} -
     (\frac{91674}{\sqrt{5}} + 40998)(\lambda_3 - \lambda_2)\,, \nonumber \\
p_{44} & = & 662860 + 296440\,{\sqrt{5}} -
   (\frac{115466}{\sqrt{5}} + 51638)(\lambda_3 - \lambda_2)\,. \nonumber
\end{eqnarray}

For two nearest-neighbor sites $A_1$ and $A_2$ separated by the distance
$n=1$ in the SOC state the constants $p_{ij}$ have other values because
the matrix $\bf A$ is degenerate. Hence we get
\begin{eqnarray}
\label{f5.12}
P_1(1,1)=0\,, && P_1(2,2)=\frac{175-78\sqrt{5}}{16}\,, \nonumber \\
P_1(3,3)=\frac{119-47\sqrt{5}}{128}\,, &&
P_1(4,4)=\frac{87-31\sqrt{5}}{128}\,, \nonumber \\
P_1(1,2)=P_1(2,1)=\frac{5(13\sqrt{5}-29)}{32}\,, &&
P_1(2,3)=P_1(3,2)=\frac{169\sqrt{5}-369}{128}\,, \nonumber \\
P_1(1,3)=P_1(3,1)=\frac{265-117\sqrt{5}}{128}\,, &&
P_1(2,4)=P_1(4,2)=\frac{171\sqrt{5}-371}{128}\,, \nonumber \\
P_1(1,4)=P_1(4,1)=\frac{235-103\sqrt{5}}{128}\,, &&
P_1(3,4)=P_1(4,3)=\frac{81-29\sqrt{5}}{128}\,. \nonumber
\end{eqnarray}

\section{Conclusion.}

In this article, we have investigated the Abelian sandpile model on the
Husimi lattice of square plaqettes. The distribution
of height probabilities and the two-point correlation
function in the SOC state have been calculated exactly. We have
shown that correlations decay with distance as
$(\lambda_3/\lambda_2)^{-(n-1)}$.

At the end we want to compare the distribution
of height probabilities obtained on different
lattices with the same coordination number $q=4$:

\bigskip
\begin{tabular}{|c|c|c|c|c|} \hline
     &  square~\cite{MD,P} & Bethe~\cite{DM} & Husimi of 
triangles~\cite{PS} & Husimi of squares \\[2mm] \hline
P(1) &  0.07363         & 0.07407        & 0.07031     
        & 0.07377 \\[2mm]
P(2) &  0.1739          & 0.22222        & 0.16406       
        & 0.20574 \\[2mm]
P(3) &  0.3063          & 0.33333        & 0.33594      
        & 0.33074 \\[2mm]
P(4) &  0.4461          & 0.37037        & 0.42969      
        & 0.38975 \\
\hline
\end{tabular}

\bigskip
From this table we see that the Husimi lattice is a good
approximation for the regular lattices and the best one is achieved
for $P(1)$ as it can be considered as a local structural
characteristic of the model~\cite{P}.
The next step of our investigations will be the calculation
of  dynamic characteristics of ASM, e.g., critical exponents
of avalanches. The choice of the Husimi lattice
gives us hope to find the relationship between the chaos and SOC state
since some spin models formulated on this lattice show the chaotic
behavior~\cite{Mon,NLO}.

\section*{Acknowledgments}

We are grateful to N.S.~Ananikian and V.B.~Priezzhev
for fruitful discussions.

One of us (R.R.S.) was supported by German
Bundesministerium f\"ur Fors\-chung and Technologie under the
grant N 211-5291 YPI and by Royal Swedish Academy of Sciences under
the research program of International Center for Fundamental
Physics in Moscow.


\newpage
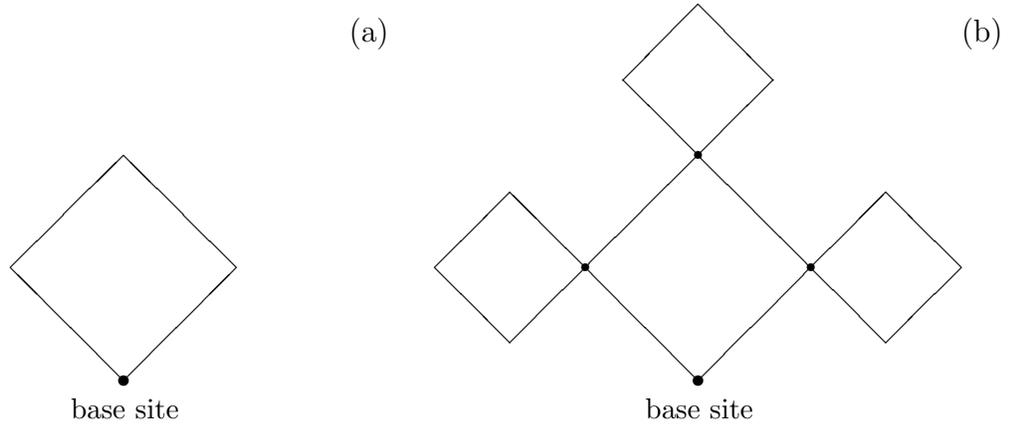
\begin{figure}[p]
\unitlength 1mm
\begin{picture}(75,60)
\put(65,55){\mbox{(a)}}
%
%
%
\put(35,10){\line(-1,1){15}}
\put(35,10){\line(1,1){15}}
\put(20,25){\line(1,1){15}}
\put(50,25){\line(-1,1){15}}
\put(35,10){\circle*{1.5}}
\put(28,5){\mbox{\small base site}}
%
\end{picture}
%
\begin{picture}(75,75)
%
%
%
\put(35,10){\line(-1,1){25}}
\put(35,10){\line(1,1){25}}
\put(10,15){\line(1,1){35}}
\put(60,15){\line(-1,1){35}}
\put(0,25){\line(1,1){10}}
\put(0,25){\line(1,-1){10}}
\put(35,60){\line(-1,-1){10}}
\put(35,60){\line(1,-1){10}}
\put(70,25){\line(-1,1){10}}
\put(70,25){\line(-1,-1){10}}
\put(20,25){\circle*{1}}
\put(35,40){\circle*{1}}
\put(50,25){\circle*{1}}
\put(35,10){\circle*{1.5}}
\put(28,5){\mbox{\small base site}}
%
\put(70,55){\mbox{(b)}}
\end{picture}
\caption{(a) A first-generation branch consists of a single
square plaquette.
(b) A second-generation branch.}
\end{figure}

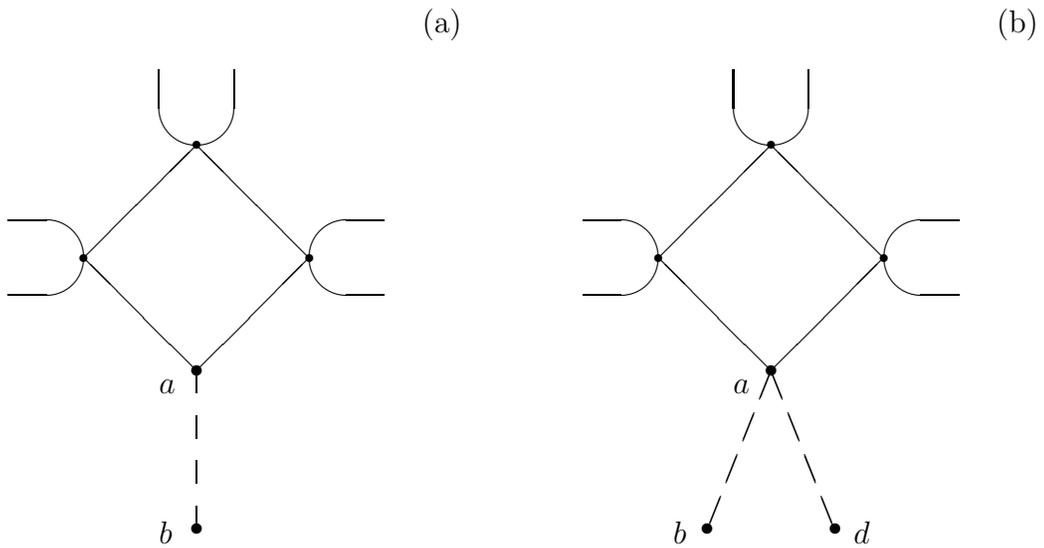
\begin{figure}[p]
\unitlength 1mm
\begin{picture}(75,75)
\put(65,70){\mbox{(a)}}
%
%
%
\put(35,25){\line(-1,1){15}}
\put(35,25){\line(1,1){15}}
\put(20,40){\line(1,1){15}}
\put(50,40){\line(-1,1){15}}
%
%
\put(10,40){\oval(20,10)[r]}
%
%
\put(35,65){\oval(10,20)[b]}
%
%
\put(60,40){\oval(20,10)[l]}
\put(20,40){\circle*{1}}
\put(35,55){\circle*{1}}
\put(50,40){\circle*{1}}
\put(35,25){\circle*{1.5}}
\put(30,22){$a$}
\multiput(35,25)(0,-6){4}{\line(0,-1){3}}
\put(35,4){\circle*{1.5}}
\put(30,2){$b$}
%
\end{picture}
%
\begin{picture}(75,75)
\put(65,70){\mbox{(b)}}
%
%
%
\put(35,25){\line(-1,1){15}}
\put(35,25){\line(1,1){15}}
\put(20,40){\line(1,1){15}}
\put(50,40){\line(-1,1){15}}
%
%
\put(10,40){\oval(20,10)[r]}
%
%
\put(35,65){\oval(10,20)[b]}
%
%
\put(60,40){\oval(20,10)[l]}
\put(20,40){\circle*{1}}
\put(35,55){\circle*{1}}
\put(50,40){\circle*{1}}
\put(35,25){\circle*{1.5}}
\put(30,22){$a$}
\multiput(35,25)(-2.228,-5.571){4}{\line(-2,-5){1.5}}
\put(26.5,4){\circle*{1.5}}
\put(22,2){$b$}
\multiput(35,25)(2.228,-5.571){4}{\line(2,-5){1.5}}
\put(43.5,4){\circle*{1.5}}
\put(46,2){$d$}
%
\end{picture}
\caption{(a) An $n$th-generation branch $G_n$ and vertex $b$ form a
subgraph $G'$. (b) Now two vertices $b$ and $d$ and the $G_n$
form a subgraph $G''$.}
\end{figure}

\begin{figure}[p]
\begin{center}
\unitlength 1mm
\begin{picture}(75,60)
%
%
%
\put(35,15){\line(-1,1){15}}
\put(35,15){\line(1,1){15}}
\put(20,30){\line(1,1){15}}
\put(50,30){\line(-1,1){15}}
%
%
\put(10,30){\oval(20,15)[r]}
\put(7,29){$G_{n-1}^{(1)}$}
%
%
\put(35,55){\oval(15,20)[b]}
\put(31,53){$G_{n-1}^{(2)}$}
%
%
\put(60,30){\oval(20,15)[l]}
\put(54,29){$G_{n-1}^{(3)}$}
\put(20,30){\circle*{1.5}}
\put(23,29){$a_1$}
\put(35,45){\circle*{1.5}}
\put(33.5,40){$a_2$}
\put(50,30){\circle*{1.5}}
\put(43,29){$a_3$}
\put(35,15){\circle*{1.5}}
\put(34,10){$a$}
\end{picture}
\end{center}
\caption{The $n$th-generation branch $G_n$ with
three nearest $(n-1)$th-generation branches $G_{n-1}^{(1)}$,
$G_{n-1}^{(2)}$ and $G_{n-1}^{(3)}$.}
\end{figure}
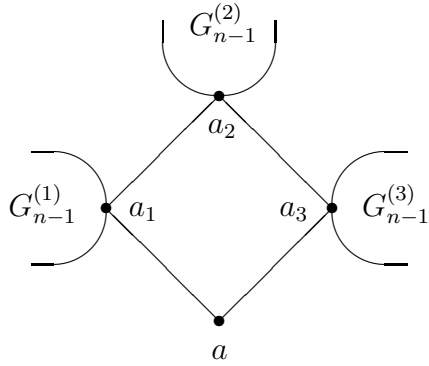

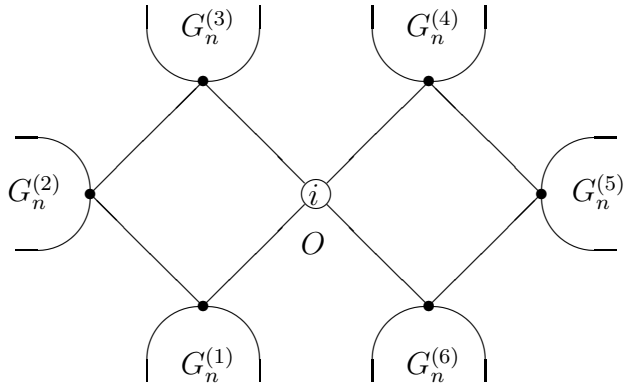
\begin{figure}[p]
\begin{center}
\unitlength 1mm
\begin{picture}(100,75)
%
%
%
\put(35,20){\line(-1,1){15}}
\put(35,20){\line(1,1){13.5}}
\put(20,35){\line(1,1){15}}
\put(35,50){\line(1,-1){13.5}}
%
%
\put(35,10){\oval(15,20)[t]}
\put(32,11){$G_{n}^{(1)}$}
%
%
\put(10,35){\oval(20,15)[r]}
\put(9,34){$G_{n}^{(2)}$}
%
%
\put(35,60){\oval(15,20)[b]}
\put(32,56){$G_{n}^{(3)}$}
\put(20,35){\circle*{1.5}}
\put(35,50){\circle*{1.5}}
\put(35,20){\circle*{1.5}}
%
%
%
\put(65,20){\line(-1,1){13.5}}
\put(65,20){\line(1,1){15}}
\put(65,50){\line(-1,-1){13.5}}
\put(80,35){\line(-1,1){15}}
%
%
\put(65,60){\oval(15,20)[b]}
\put(62,56){$G_{n}^{(4)}$}
%
%
\put(90,35){\oval(20,15)[l]}
\put(84,34){$G_{n}^{(5)}$}
%
%
\put(65,10){\oval(15,20)[t]}
\put(62,11){$G_{n}^{(6)}$}
\put(65,50){\circle*{1.5}}
\put(80,35){\circle*{1.5}}
\put(65,20){\circle*{1.5}}
\put(50,35){\circle{4}}
\put(48,27){$O$}
\put(49,33.5){$i$}
\end{picture}

\end{center}
\caption{A site $O$ with height $i$ is located deep inside the lattice
and surrounded by the six $n$th-generation branches
$G_n^{(1)}, G_n^{(2)}, G_n^{(3)}, G_n^{(4)}, G_n^{(5)}$ and $G_n^{(6)}$.}
\end{figure}

\begin{figure}[p]
\begin{center}
\unitlength 1mm
\begin{picture}(160,60)
\put(10,30){\line(1,0){40}}
\multiput(50,30)(4,0){10}{\line(1,0){2}}
\put(90,30){\line(1,0){60}}
\put(10,30){\line(0,-1){20}}
\put(10,10){\line(1,0){20}}
\put(30,10){\line(0,1){40}}
\put(30,50){\line(1,0){20}}
\put(50,50){\line(0,-1){20}}
\put(40,0){\oval(24,24)[lt]}
\put(32,2){$U_{1}^{(b)}$}        
\put(30,10){\circle*{1.5}}
\put(0,0){\oval(24,24)[rt]}
\put(1,2){$U_{0}^{(b)}$}         
\put(10,10){\circle*{1.5}}
\put(0,40){\oval(24,24)[rb]}
\put(2,33){$U_{0}$}              
\put(10,30){\circle*{1.5}}
\put(20,60){\oval(24,24)[rb]}
\put(22,54){$U_{1}^{(t)}$}       
\put(30,50){\circle*{1.5}}
\put(60,60){\oval(24,24)[lb]}
\put(53,54){$U_{2}^{(t)}$}       
\put(50,50){\circle*{1.5}}
\put(46,24){$A_2$}
\put(50,30){\circle*{1.5}}
\put(24,33){$A_1$}
\put(30,30){\circle*{2}}
\put(90,30){\line(0,-1){20}}
\put(90,10){\line(1,0){20}}
\put(110,10){\line(0,1){40}}
\put(110,50){\line(1,0){20}}
\put(130,50){\line(0,-1){40}}
\put(130,10){\line(1,0){20}}
\put(150,10){\line(0,1){20}}
\put(100,60){\oval(24,24)[rb]}
\put(102,54){$U_{n}^{(t)}$}       
\put(110,50){\circle*{1.5}}
\put(140,60){\oval(24,24)[lb]}
\put(132,54){$U_{n+1}^{(t)}$}       
\put(130,50){\circle*{1.5}}
\put(160,40){\oval(24,24)[lb]}
\put(152,33){$U_{n+2}$}              
\put(150,30){\circle*{1.5}}
\put(160,0){\oval(24,24)[lt]}
\put(152,2){$U_{n+2}^{(b)}$}        
\put(150,10){\circle*{1.5}}
\put(130,0){\oval(10,20)[t]}
\put(126,2){$U_{n+1}^{(b)}$}         
\put(130,10){\circle*{1.5}}
\put(110,0){\oval(10,20)[t]}
\put(107,2){$U_{n}^{(b)}$}        
\put(110,10){\circle*{1.5}}
\put(80,0){\oval(24,24)[rt]}
\put(81,2){$U_{n-1}^{(b)}$}         
\put(90,10){\circle*{1.5}}
\put(89,33){$A_{n-1}$}
\put(90,30){\circle*{1.5}}
\put(113,24){$A_n$}
\put(110,30){\circle*{1.5}}
\put(133,33){$A_{n+1}$}
\put(130,30){\circle*{2}}
\end{picture}
\end{center}
\caption{The path from the site $A_1$ to the site $A_{n+1}$
on the Husimi lattice goes through the points $A_2$,\ldots, $A_n$.
The left-hand side of the lattice beginning from the vertex $A_k$,
$k= 1,\ldots,n$, is denoted as a branch $G_k$ with the root$A_k$.}
\end{figure}
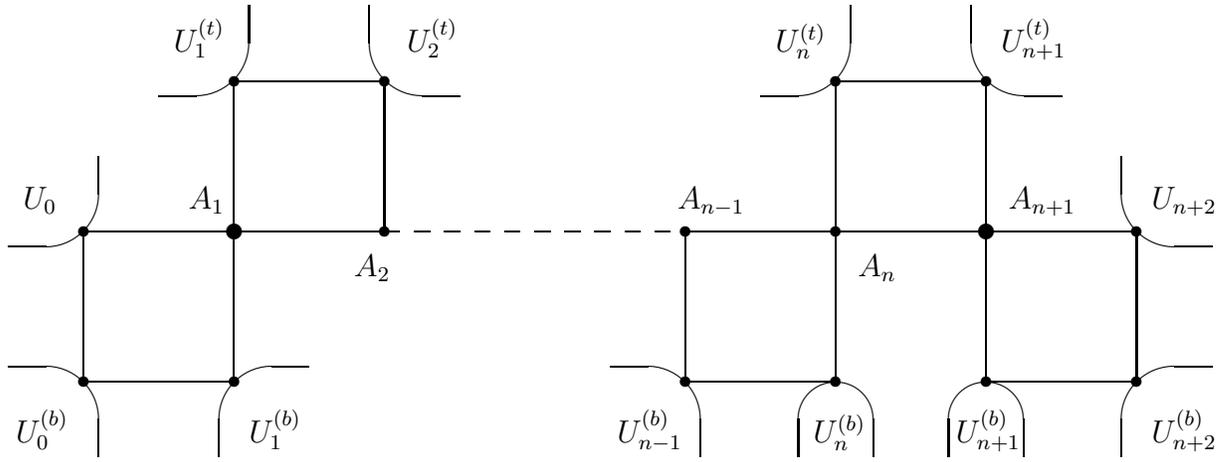

\end{document}